\documentclass[aps,
longbibliography,
twocolumn,superscriptaddress,
showpacs]{revtex4-1}
\usepackage{amssymb, graphicx}
\usepackage{enumerate}
\usepackage{amsmath,bm,bbold}
\usepackage{dcolumn}
\usepackage{mathrsfs}
\usepackage{color}
\usepackage[latin1]{inputenc}
\usepackage[normalem]{ulem} 

\usepackage[version=3]{mhchem}
 
\makeatletter

\begin{document}
\title{Contribution of the pre-ionized H$_2$ and the ionized H$_2^+$ subsystems to the HHG Spectra of H$_2$ in intense laser fields
}

\author{Hossein Iravani}
\affiliation{Department of Chemistry, Tarbiat Modares University, P. O. Box 14115-175, Tehran, I. R. Iran}
\author{Hassan Sabzyan}
\email{E-mail: sabzyan@sci.ui.ac.ir}
\affiliation{Department of Chemistry, University of Isfahan, Isfahan 81746-73441, I. R. Iran}
\author{Mohsen Vafaee}
\email{E-mail: m.vafaee@modares.ac.ir}
\affiliation{Department of Chemistry, Tarbiat Modares University, P. O. Box 14115-175, Tehran, I. R. Iran}
\author{Behnaz Buzari}
\affiliation{Department of Chemistry, University of Isfahan, Isfahan 81746-73441, I. R. Iran}

\begin{abstract}
 Contributions of the pre-ionized H$_2$ (PI-H$_2$) and ionized H$_2^+$ subsystems of the two-electron H$_2$ system to its high-order harmonic generation in 8-cycle $\sin^{2}$-like ultrafast intense laser pulses are calculated and analyzed based on the solution of the time-dependent Schr\"{o}dinger equation (TDSE) for the one-dimensional two-electronic H$_2$ system with fixed nuclei. The laser pulses have $\lambda\ =\ 390\ \&\ 532$ nm wavelengths and $I=1\times 10^{14}$, 5$\times 10^{14}$, 1$\times 10^{15}$ \&\ 5$\times 10^{15}$ Wcm$^{-2}$ intensities. 
It is found that at the two lower intensities, the PI-H$_2$ subsystem dominantly produces the HHG spectra. While, at the two higher intensities, both PI-H$_2$ and ionized H$_2^+$ subsystems contribute comparably to the HHG spectra. In the H$_2^+$ subsystem, the symmetry of the populations of H$_2^+$(I) and H$_2^+$(II) regions (left and right regions of H$_2^+$ subsystem) is broken by increasing the laser intensity. 
Complex patterns and even harmonics also appear at these two higher intensities. For instance, at $1\times 10^{15}$ Wcm$^{-2}$ intensity and $\lambda$ = 532 nm wavelength, the even harmonics are appeared near cut-off region. Interestingly, at $5\times 10^{15}$ Wcm$^{-2}$ intensity and $\lambda$ = 390 nm wavelength, the even harmonics replaced by the odd harmonics with red shift.
At $\lambda$ = 390 \&\ 532 nm wavelengths and $I=1\times 10^{15}$ intensity, the two-electron cutoffs corresponding to nonsequential double-recombination (NSDR) with maximum return kinetic energy of 4.70$U_p$ are detected. 
The HHG spectra of the whole H$_2$ system obtained with and without nuclear dynamics treated classically are approximately similar. However, at $1\times 10^{15}$ Wcm$^{-2}$ intensity and $\lambda$ = 532 nm wavelength, if we take into account nuclear dynamics, the even harmonics which are appeared near cutoff region, replaced by the odd harmonics with blue shift.

\end{abstract}

\maketitle

\section{Introduction}

The HHG and related processes such as above-threshold ionization (ATI) and dissociation, bond hardening and softening produced by atomic and molecular systems in strong laser fields have received great attentions in the past two decades [1-7]. During these years, HHG process has been an important subject in both experimental and theoretical physics [8,9].
A molecular system exposed to a strong laser field, display different phenomena like one-electron ionization, multi-electron ionization and dissociation.
The three step model has proven very successful as a basis to explain the atomic HHG and molecular aspects of HHG not present in atoms [8].  

In spite of recent progresses in computing facilities, direct solution of the time-dependent Schr\"{o}dinger equation (TDSE) in three-dimension (3D) can only be carried out for one-electron systems due to limitations of the hardware facilities and time.
Krause et al. were the first to solve the 3-D time dependent Schr\"{o}dinger equation (TDSE) for one-electron systems such as He$^+$ and calculate the harmonic spectrum produced by them [11]. 
To facilitate the numerical solution of TDSE, many studies have been done in lower dimensions (1D and 2D) [13-16] in which the electron-nuclear Coulomb interaction modeled by so-called soft-core Coulomb potentials [17-19].

\begin{figure}[ht]
\begin{center}
\begin{tabular}{c}
\resizebox{90mm}{!}{\includegraphics{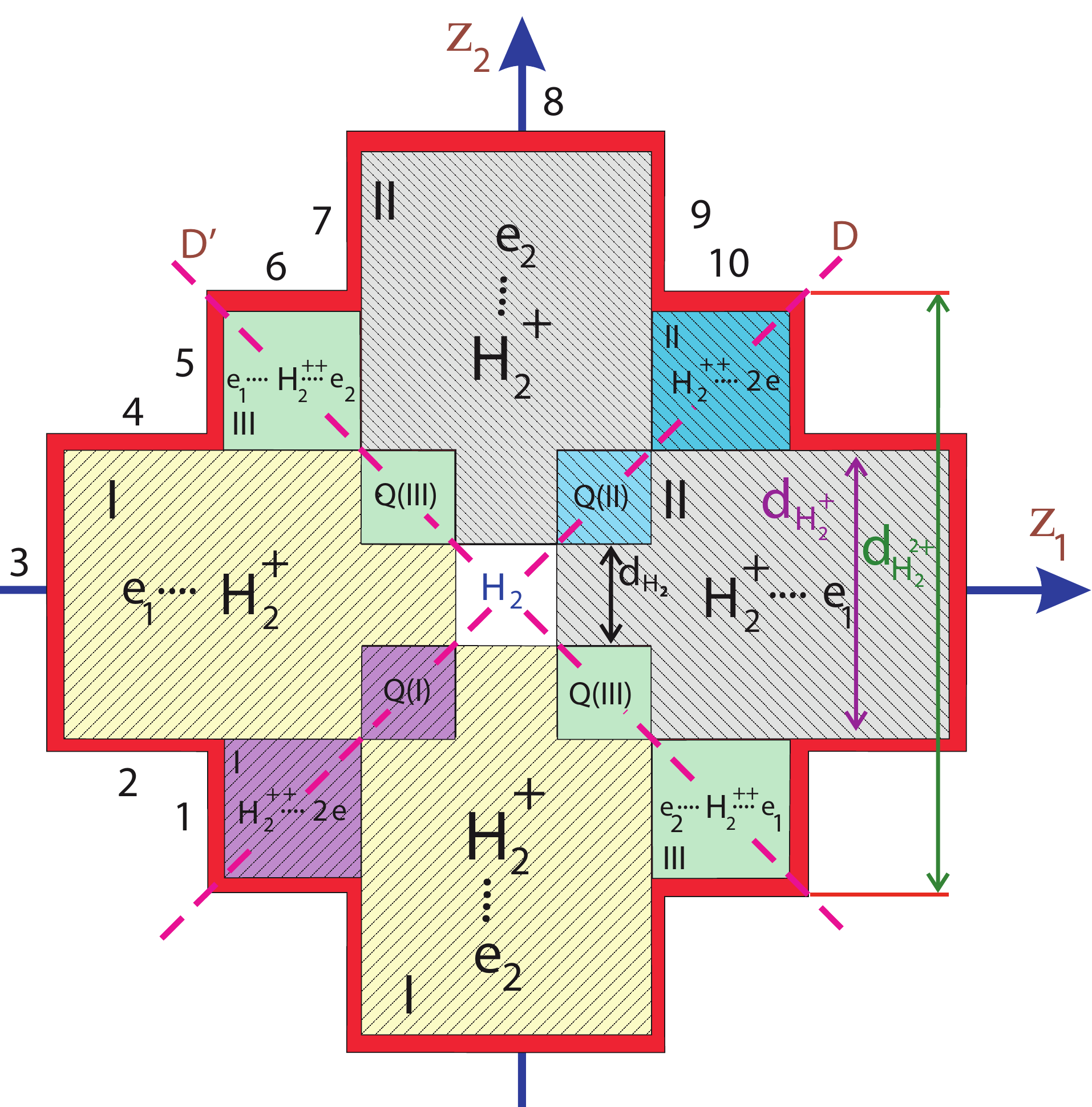}}
\end{tabular}
\caption{
\label{box}
The schematic diagram of the simulation box adopted in this study. Roman and Arabic numbers represent the box partitions and their borders, respectively. Reprinted from [29], with the permission of AIP Publishing.
		}
\end{center}
\end{figure}

\begin{figure*}[ht]
\begin{center}
\begin{tabular}{c}
\resizebox{120mm}{!}{\includegraphics{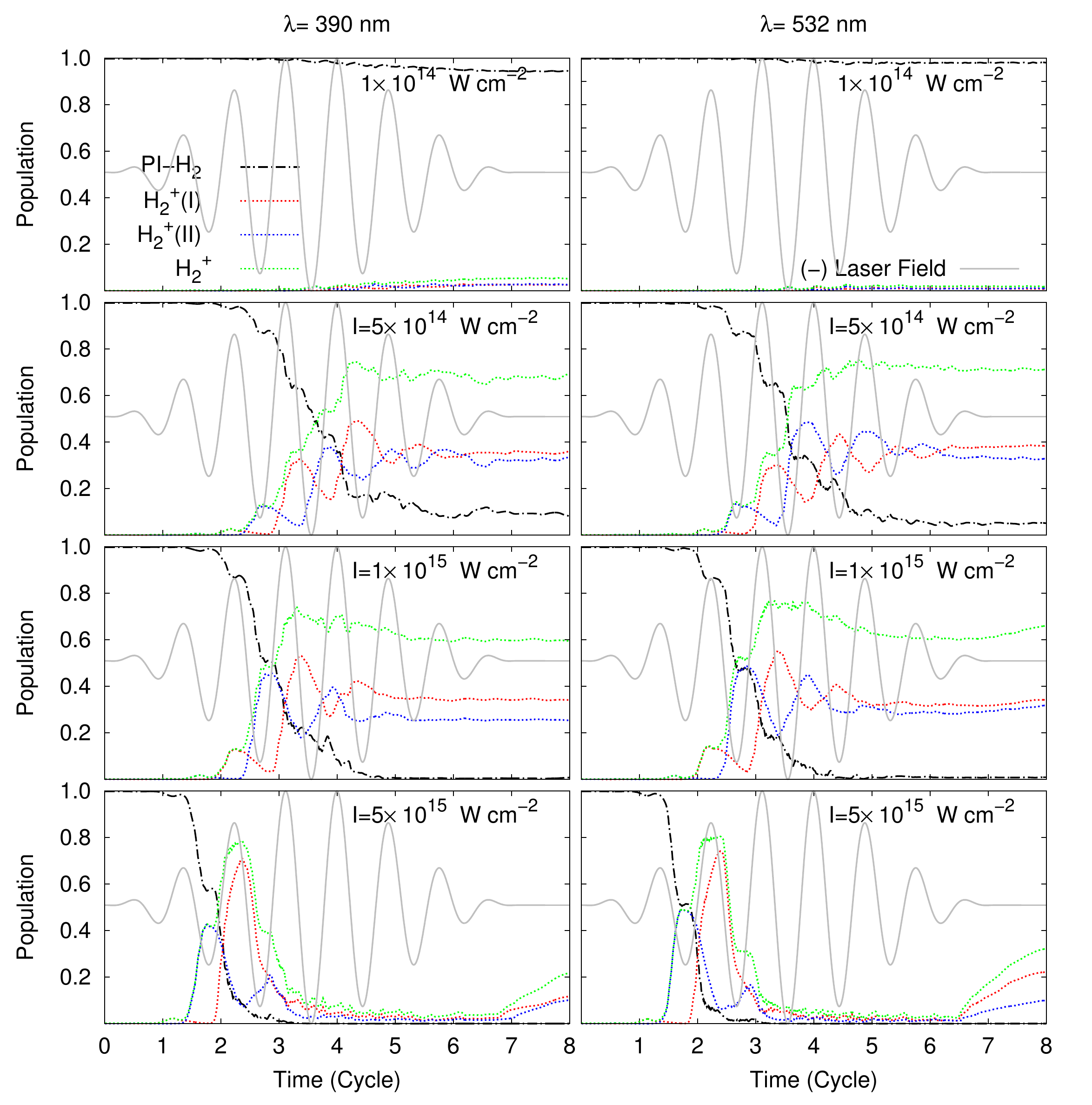}}
\end{tabular}
\caption{
\label{pop}
Variations of the populations of the pre-ionized-H$_2$ (PI-H$_2$), ionized H$_2^+$, and two regions of H$_2^+$ subsystem (i.e. H$_2^+$(I) and H$_2^+$(II)), introduced in Fig. 1, during the interaction of the H$_2$ system with 8-cycle $\sin^{2}$-like ultrafast intense laser pulses of $\lambda$ = 390 \&\ 532 nm wavelengths and at $I=1\times 10^{14}, 5\times10^{14}$, $1\times 10^{15}$ \&\ $5\times 10^{15}$ Wcm$^{-2}$ intensities.
		}
\end{center}
\end{figure*}


\begin{figure*}[ht]
\begin{center}
\begin{tabular}{c}
\resizebox{120mm}{!}{\includegraphics{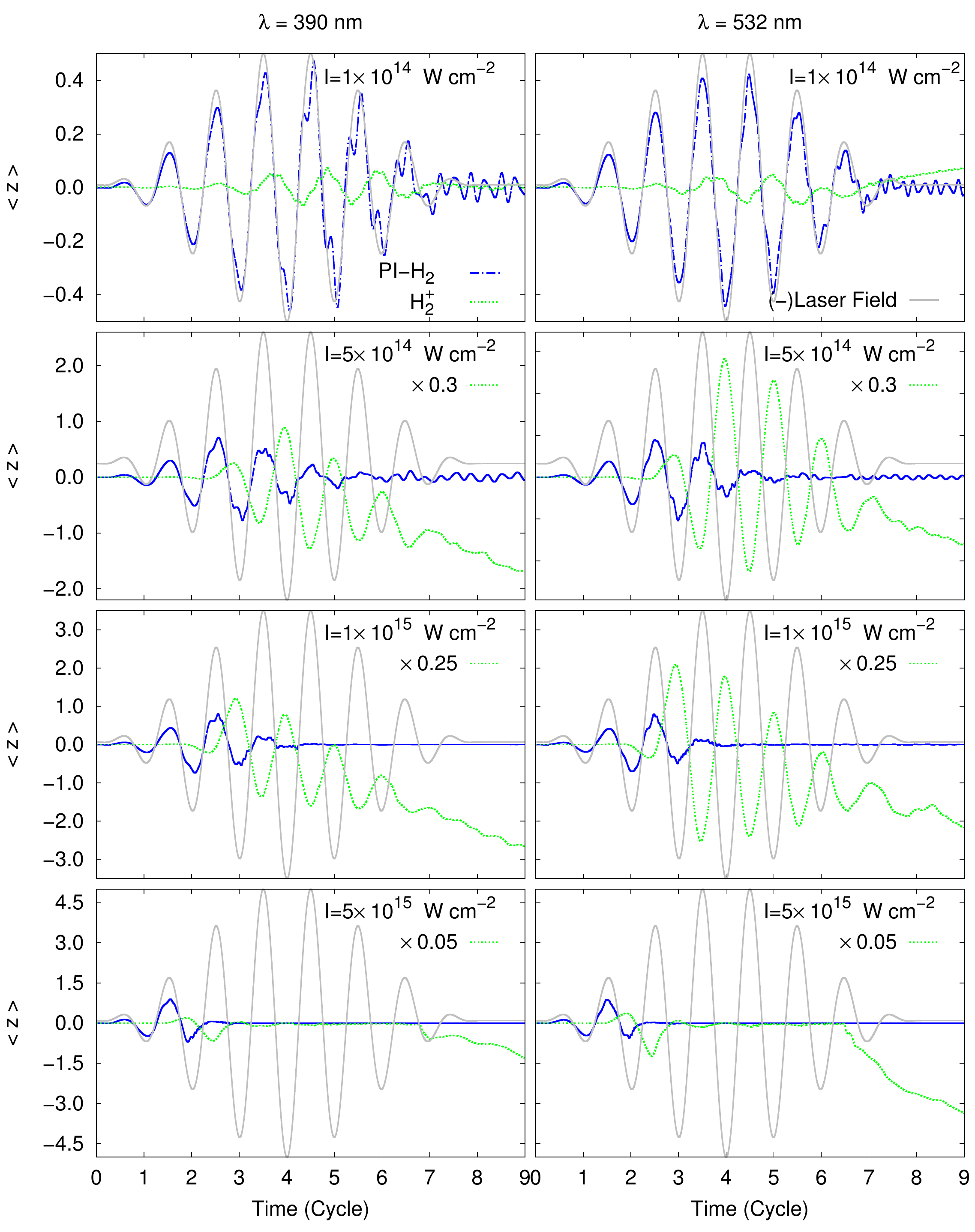}}
\end{tabular}
\caption{
\label{mom}
Variations of the expectation (average) values of electron position, $\langle z\rangle$, calculated for the PI-H$_2$ and H$_2^+$ subsystems of the H$_2$ system during its interaction with an 8-cycle $\sin^{2}$-like ultrafast intense laser pulses of $\lambda$ = 390 \&\ 532 nm wavelengths and  $I=1\times10^{14}, 5\times10^{14}, 1\times10^{15}$ \&\ $5\times10^{15}$ Wcm$^{-2}$ intensities. See Fig. 1 for partitioning scheme.
		}
\end{center}
\end{figure*}

\begin{figure*}[ht]
\begin{center}
\begin{tabular}{c}
\resizebox{120mm}{!}{\includegraphics{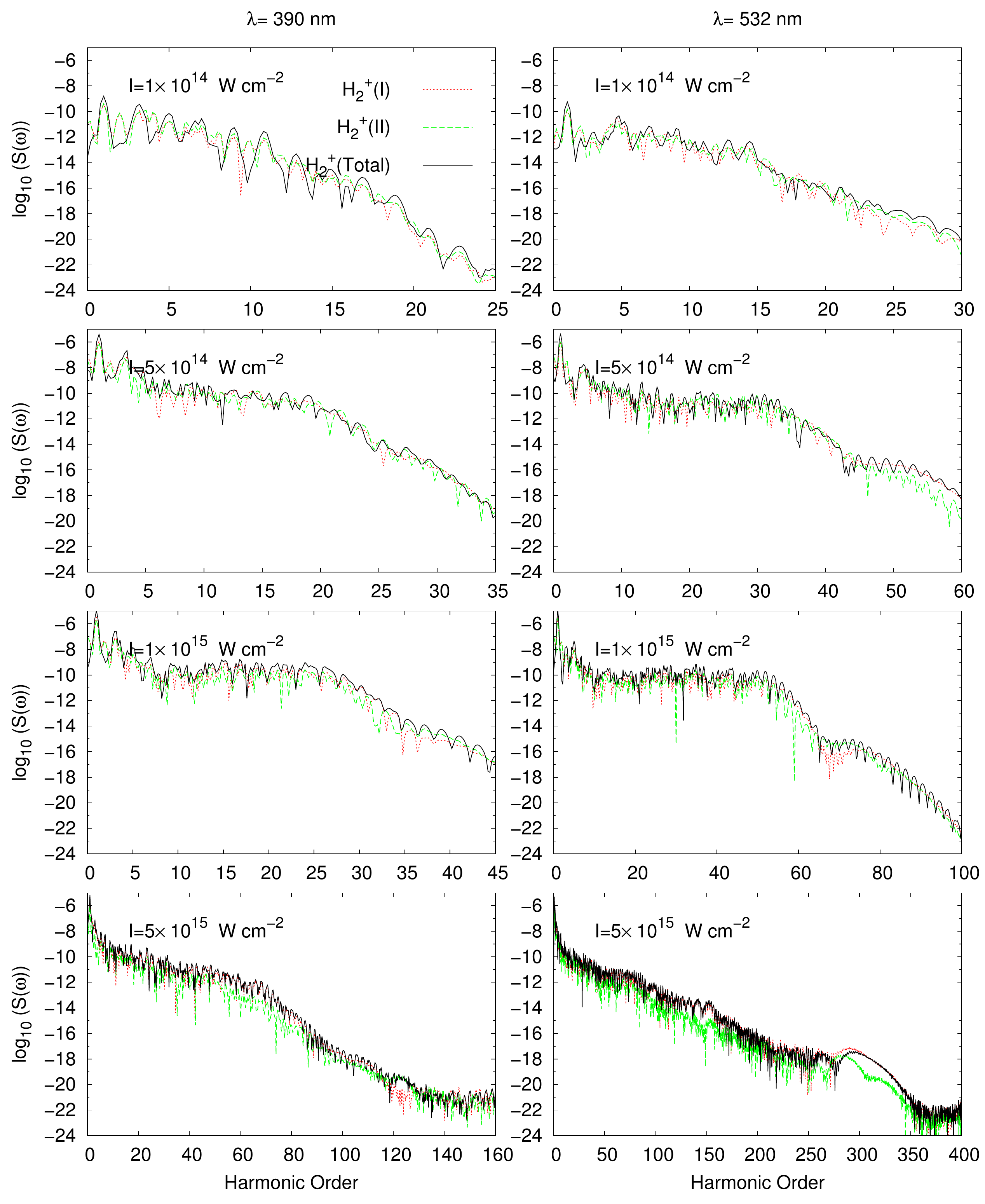}}
\end{tabular}
\caption{
\label{hhg-ionic_h2+}
The HHG spectra, $S(\omega)$, produced by the two regions of H$_2^+$ subsystem (i.e. H$_2^+$(I) and H$_2^+$(II)), as defined in Fig. 1, during the interaction with 8-cycle $\sin^{2}$-like ultrafast intense laser pulses of $\lambda$ = 390 \&\ 532 nm wavelengths and $I=1\times 10^{14}, 5\times10^{14}$, $1\times 10^{15}$ \&\ $5\times 10^{15}$ W cm$^{-2}$ intensities. Note that different scales are used for different plots.
    		}
\end{center}
\end{figure*}

\begin{figure*}[ht]
\begin{center}
\begin{tabular}{c}
\resizebox{120mm}{!}{\includegraphics{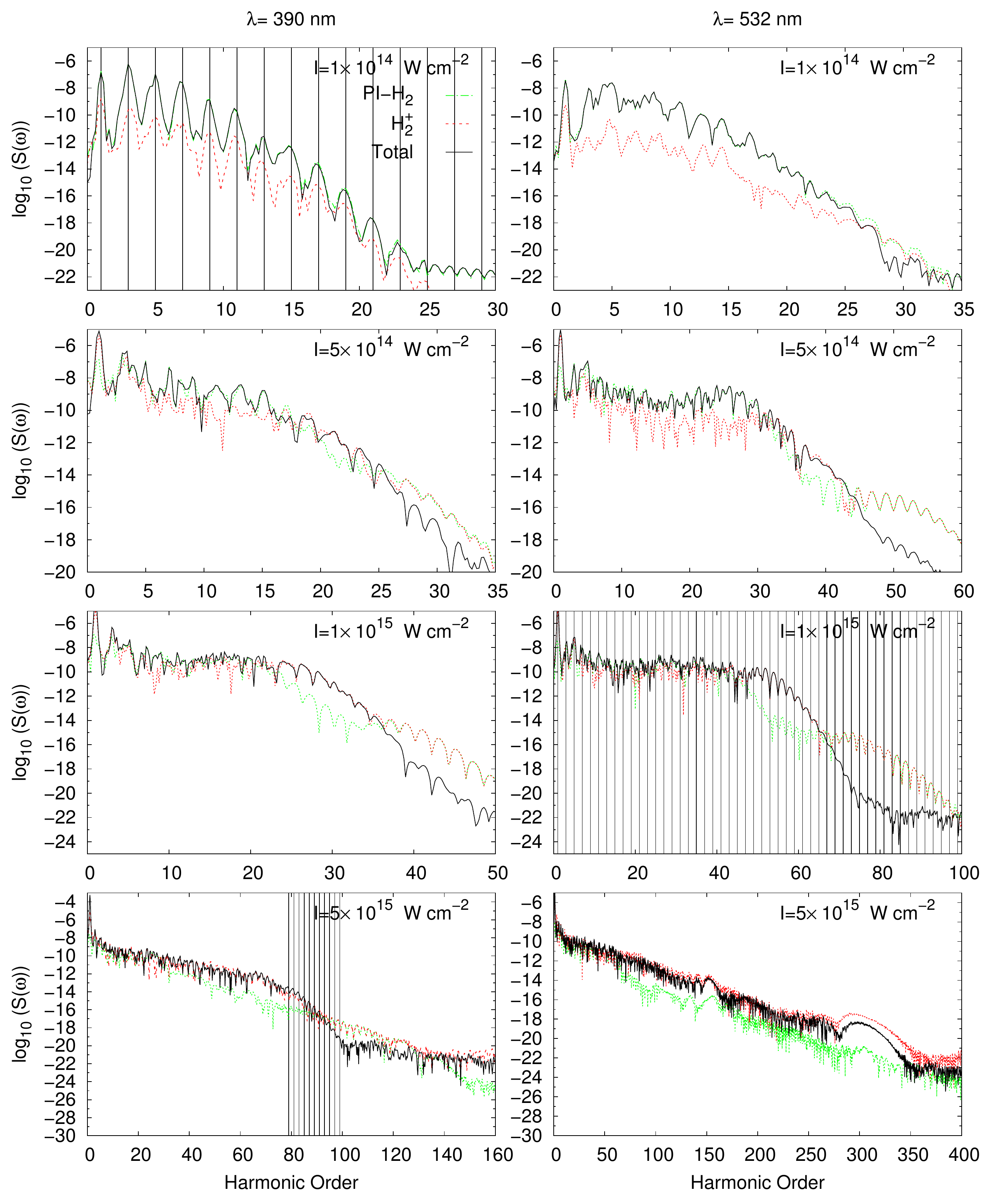}}
\end{tabular}
\caption{
\label{hhg_tot}
The same as Fig. 4, but for the PI-H$_2$ and H$_2^+$ subsystems and the whole H$_2$ system (see Fig. 1). The vertical lines shown in some plots denote the odd harmonic orders. 
}
\end{center}
\end{figure*}

\begin{figure*}[ht]
\begin{center}
\begin{tabular}{c}
\resizebox{120mm}{!}{\includegraphics{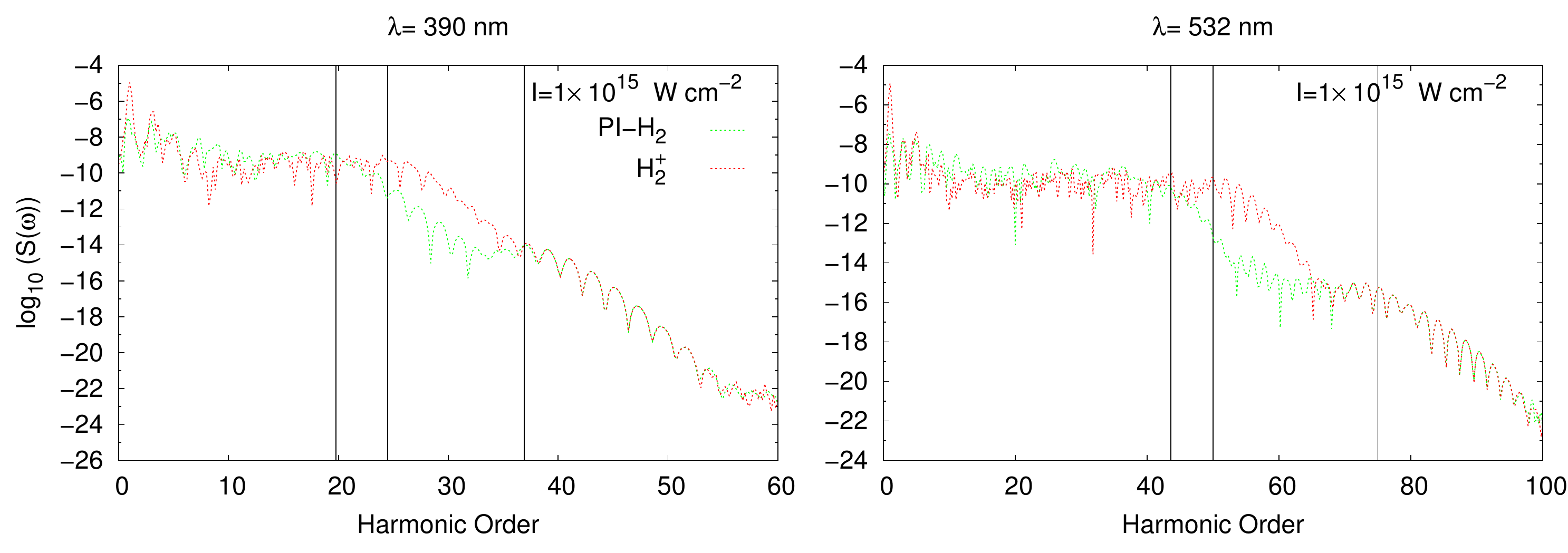}}
\end{tabular}
\caption{
\label{Madson_Cutoff}
The HHG spectra, $S(\omega)$, produced by the PI-H$_2$ and H$_2^+$ subsystems during the interaction with 8-cycle $\sin^{2}$-like ultrafast intense laser pulses of $\lambda$ = 390 \&\ 532 nm wavelengths and $I=1\times 10^{15}$ intensity. The vertical lines shown denote the one-electron cutoff of PI-H$_2$ subsystem, the one-electron cutoff of H$_2^+$ subsystem and two-electron cutoff of PI-H$_2$ subsystem respectively. 
}
\end{center}
\end{figure*}

\begin{figure*}[ht]
\begin{center}
\begin{tabular}{c}
\resizebox{180mm}{!}{\includegraphics{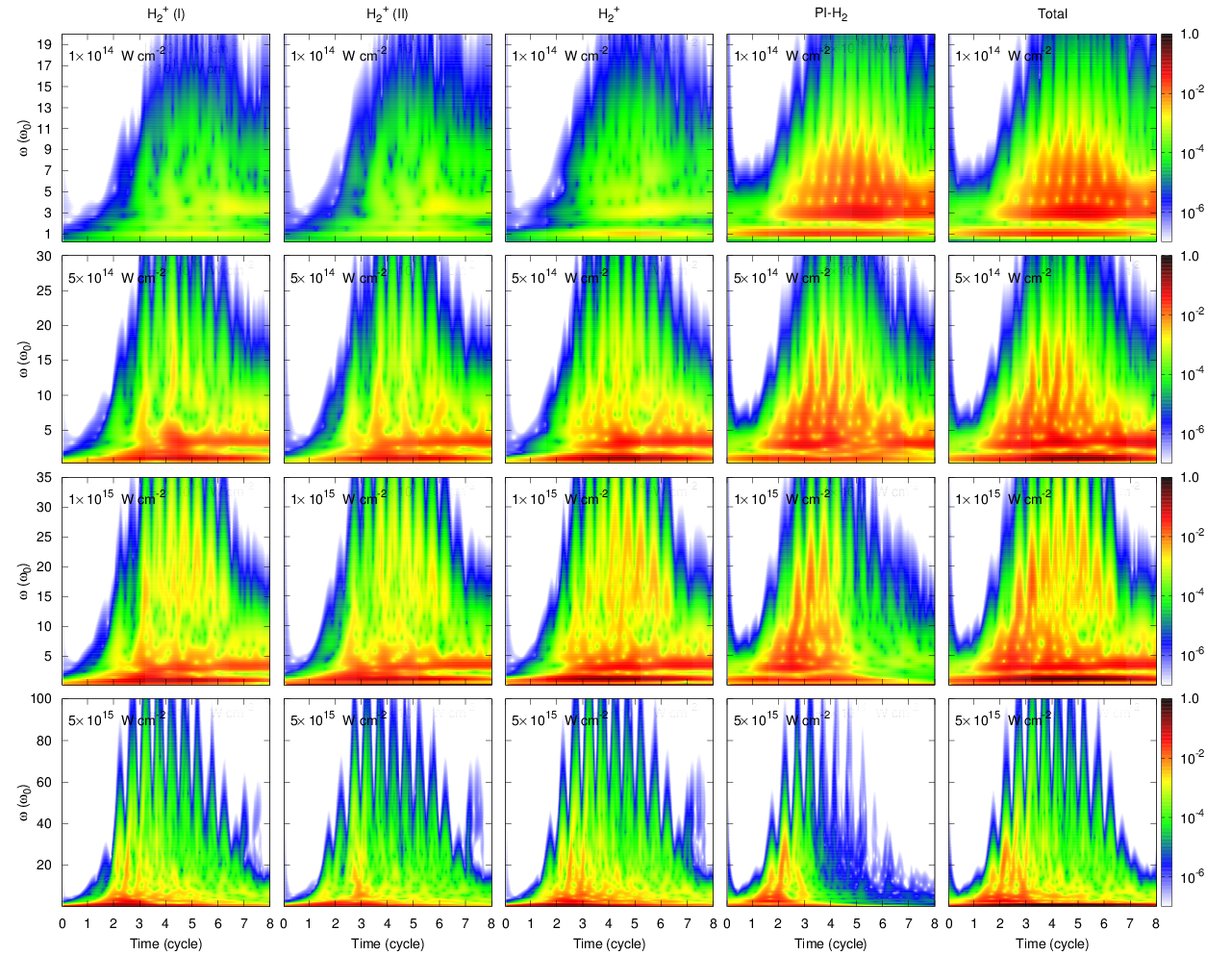}}
\end{tabular}
\caption{
\label{Mw-H2_H2+_Tot_390_and_Ioinc}
  The TFP of the HHG produced by the two-electron wave packet evolution of the pre-ionized molecular H$_2$ (PI-H$_2$) and the singly ionized H$_2^+$ subsystems (and its components H$_2^+$(I) and H$_2^+$(II)) and the overall H$_2$ system during the interaction of the H$_2$ system with 8-cycle $\sin^{2}$-like ultrafast intense laser pulses of $\lambda$ = 390 nm wavelength and at $I=1\times 10^{14}, 5\times10^{14}$, $1\times 10^{15}$ \&\ $5\times 10^{15}$ Wcm$^{-2}$ intensities.  
		}
\end{center}
\end{figure*}

\begin{figure*}[ht]
\begin{center}
\begin{tabular}{c}
\resizebox{180mm}{!}{\includegraphics{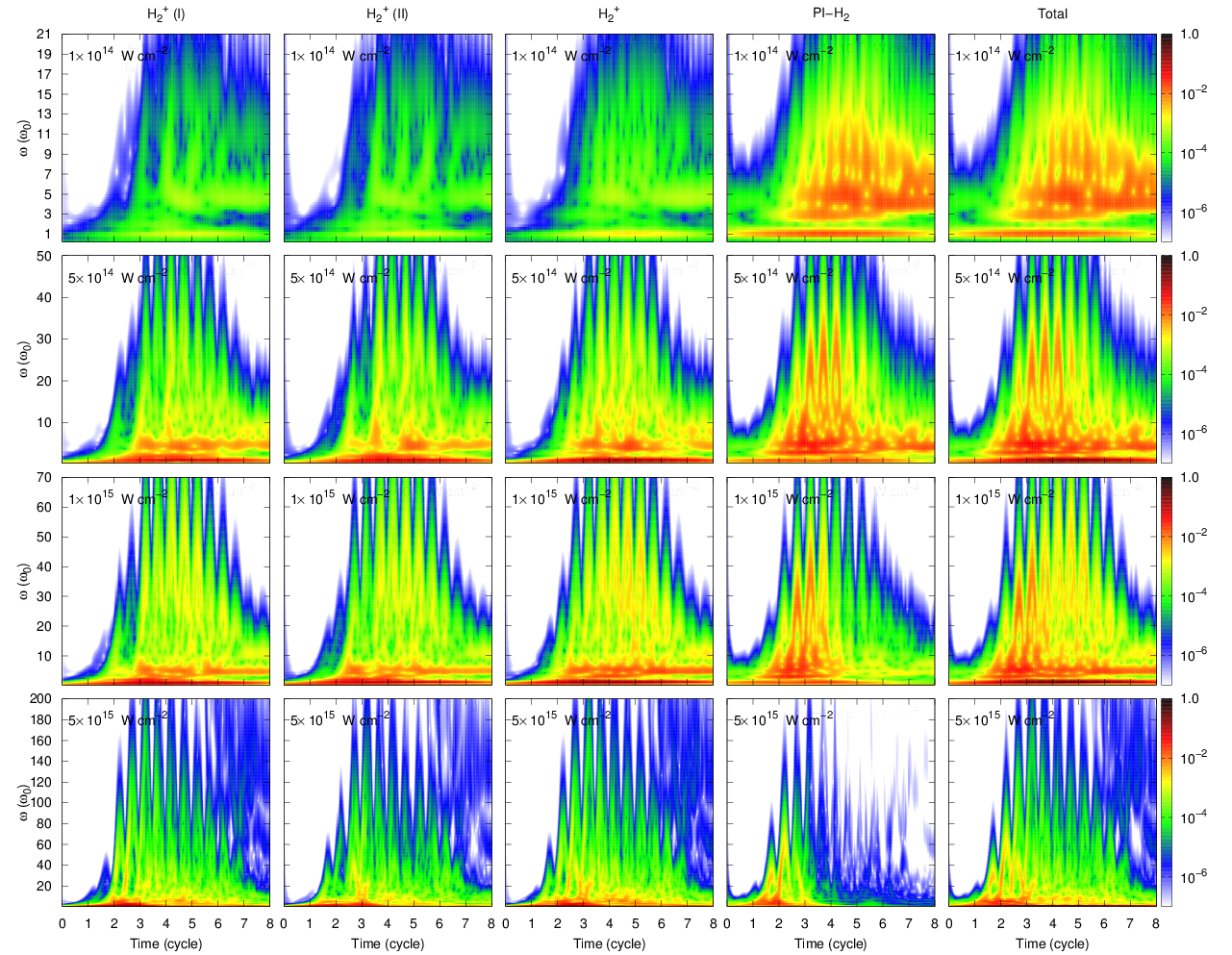}}
\end{tabular}
\caption{
\label{Mw-H2_H2+_Tot_532_and_Ioinc}
The same as Fig. 7, but for $\lambda$ =532 nm wavelength. 
		}
\end{center}
\end{figure*}

\begin{figure*}[ht]
\begin{center}
\begin{tabular}{c}
\resizebox{120mm}{!}{\includegraphics{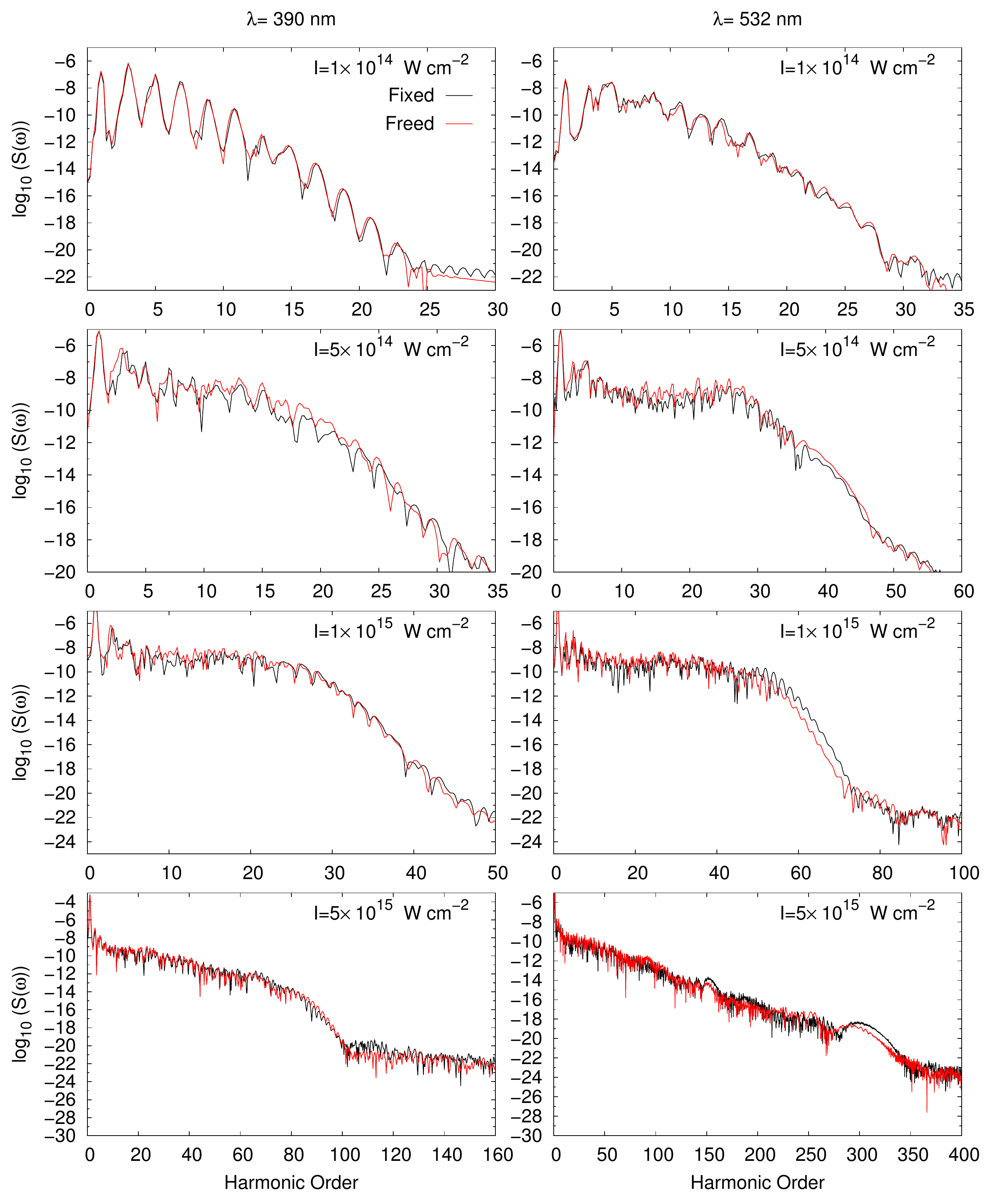}}
\end{tabular}
\caption{
\label{HHG_Tot_Moved_Fixed}
Comparing the HHG of the whole H$_2$ system (as defined in Fig. 1) for fixed and freed nuclei. 
}
\end{center}
\end{figure*}

\begin{figure*}[ht]
\begin{center}
\begin{tabular}{c}
\resizebox{120mm}{!}{\includegraphics{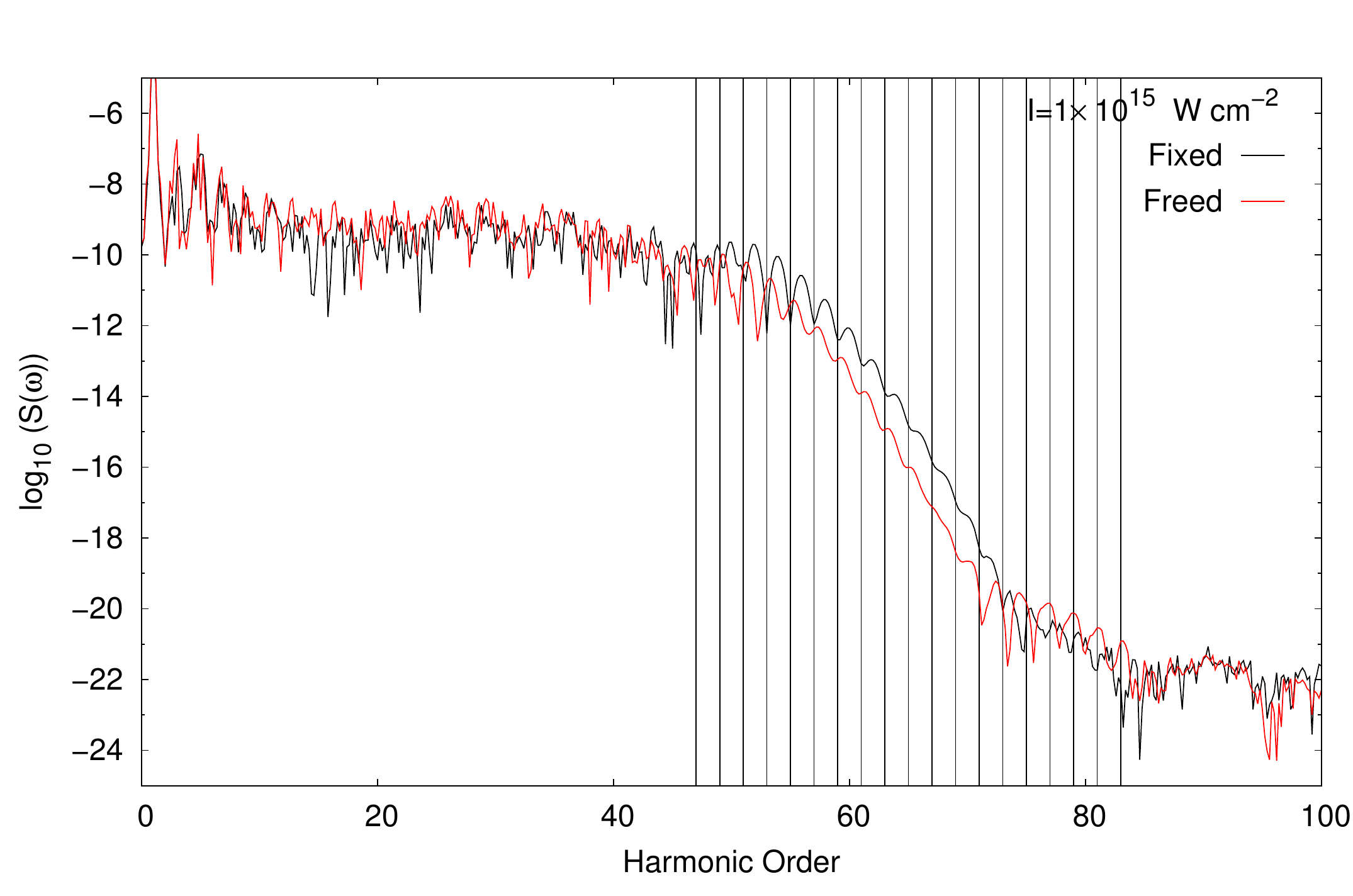}}
\end{tabular}
\caption{
\label{HHG_Tot_Moved_Fixed_odd_even}
Odd \&\ even harmonics in the HHG of the whole H$_2$ system (as defined in Fig. 1) for fixed and freed nuclei during the interaction with 8-cycle $\sin^{2}$-like ultrafast intense laser pulses of $\lambda$ =  532 nm wavelength and $I=1\times 10^{15}$ intensity. The vertical lines denote the odd harmonic orders. 
}
\end{center}
\end{figure*}

For multi-electron systems, however, using single active electron approximation is common [10]. 
In 1992, Krause et al. solved the time-dependent Hartree-Fock equations for He and the other rare gases using a single-active-electron approximation in which only one electron interacts with the field, while the others remain fixed in their ground-state orbitals [11]. 
HHG spectra are also hoped to be well captured by TDDFT with known exchange-correlation potentials. However, HHG spectra of helium which obtained by TDDFT may display an unphysical second plateau [12]. Recently, Bauer et al. introduced time-dependent renormalized-natural-orbital theory (TDRNOT) which can be formulated for any time-dependent two-electron system in either spin configuration. They found that TDRNOT with two natural orbitals per spin reproduces the HHG spectra of He very well [12].

HHG in the molecular system arises from evolution of the electronic wavepackets which can be partitioned into different electronic levels, different vibrational levels, different spatial regions (left and right side of the simulation box), or different components of a molecule in a laser field. 
To obtain the contributions of different vibrational and electronic states to the total HHG spectrum, it is possible to decompose the total wavefunction as a superposition of the several lowest Born-Oppenheimer electronic wavefunctions of the system and the residual part of the wavefunction including the higher excited states and electronic continuum states [20,21].
Total HHG signal can be also decomposed into different localized signals by introducing the electronic wavefunction localized on the right and the left nucleus [22]. 

Recently we have studied the contribution of different components of H$_2$ molecule in the HHG spectra. In [23, 24], we introduced the \textit{homolytic} and \textit{ionic} transient states of pre-ionized H$_2$ (PI-H$_2$) in the two-electron evolution of the 1D H$_2$-system. In these studies, the mechanism of forming transient species prior to ionization of the two-electron molecular H$_2$ system, exposed to ultrashort intense laser pulses, are investigated by solving the one-dimensional (1D) TDSE, including nuclear motion semiclassically and using the simulation box shown in Fig. 1. In this work, we will continue along the same line by focusing on the high-order harmonic generation by the two-electron and one-electron subsystems of the H$_2$ system, named pre-ionized H$_2$ (PI-H$_2$) and H$_2^+$, respectively.

We specially focus on the relative contributions of the PI-H$_2$ and the ionized H$_2^+$ species to the HHG of the H$_2$ system, and the effects of intensity and wavelength of the laser pulse on these contributions.
The H$_2$ system is the simplest model enabled us to investigate the complexity of the HHG of a multi-electron system. This system can be considered as a combination of an one-electron and a two-electron subsystems. 
In studying multi-electron systems, usually one active electron is taken into account, and the HHG of other electrons are ignored [25]. In this work, however, we check the amount of competition and the effect of both electrons. 

In 2007, Bauer et al. reported nonsequential double-recombination (NSDR) HHG process for an atomic system [26]. NSDR process is occurred when two electrons propagating independently of each other in the field and then returning simultaneously to emit the combined kinetic energy of the two electrons as HHG. This process results in a new plateau in the HHG spectrum which its cutoff energy is larger than the one-electron HHG signal [27].
Recently Madsen et al. have studied the signatures of NSDR HHG in a molecular system. They found that for small internuclear distances of R $\leqslant$ 8 a.u. the observed cutoffs in the NSDR HHG part of the spectra remain similar to the atomic case [28].
For internuclear distances of R $\geqslant$ 8 a.u., however, a characteristic signal is observed with even higher energy than that of atomic NSDR HHG. They claimed that this signal and the
associated cutoff originates from electrons which are both emitted and recombine within the same period. This same-period emission and recombination (SPEAR) allows a higher total kinetic energy of the electrons than what is allowed for an atomic system where electron-electron repulsion suppresses such a signal [27].
Madsen's group also formulate a Coulomb-corrected three-step model (CC-TSM) to understand the NSDR process. At internuclear distances of R $\simeq$ 8-9 a.u., they find a change in the NSDR process originating from a transition in the charge-transfer dynamics in the molecule. This change is not observed for one-electron HHG and is not predicted by the three-step model but comes from the proposed CC-TSM [28].
In this work, we search for the NSDR and SPEAR cutoffs in the HHG spectra.

This paper is organized as follows: the computational method is introduced in Section II, the results are presented in Section III , and finally, conclusions drawn from this work are highlighted in section IV. Throughout this paper, atomic units (a.u.), $e$=1, $\hbar=1$, $m_e$=1 are used unless stated otherwise.

 $\vspace{-1cm}$
 
\section{Computational Method}

We choose a one-dimensional (1D) model for the coordinates of each electron. Indeed, we solve the TDSE for a 1D model with a soft-core Coulomb interaction between the charged particles to investigate the electronic dynamics of the two-electron molecular H$_2$ in a strong laser field.
To do this, we adopt a partitioning scheme for the simulation box which has been applied in our previous works [23, 29]. We used an absorbing potential at the boundaries of the simulation box to avoid the electron reflections from the boundaries.
 
This simulation box is shown in Fig. 1 (Fig. 3 of Ref. [29]). In order to simplify the numerical solution of this electron-nuclear TDSE within the adiabatic approximation [30], the two nuclei coordinates are treated either fixed or considered as parameters adjusted every time steps using a classical Verlet approach [31]. 
The initial state is a singlet ground electronic state with an equilibrium internuclear distance R = 2.13 at rest.

In what follows, $R_1(t)$ and $R_2(t)$ indicate the instantaneous positions of the nuclei, $z_1$ and $z_2$ are the coordinates of electrons. The temporal evolution of electrons of this system is described by TDSE, i.e. [30,32]
\begin{eqnarray}\label{eq:1}
&&i\frac{\partial \psi(z_1,z_2,t; R_1, R_2)}{\partial t}
=
\nonumber \\
&&
H_e(z_1,z_2,t; R_1, R_2) \psi(z_1,z_2,t; R_1, R_2)
\end{eqnarray}
where the electronic Hamiltonian for this system, $H_e(z_1,z_2,t; R_1, R_2)$, is given by
\begin{eqnarray}\label{eq:2}
&&H_e(z_1,z_2,t; R_1, R_2) =
\nonumber \\
&&-\frac{1}{2} \left[\frac{\partial ^2}{\partial z_1^2}+\frac{\partial ^2}{\partial z_2^2}\right]
+V(z_1,z_2,t; R_1, R_2).
\end{eqnarray}
The potential $V(z_1,z_2,t; R_1, R_2)$ is given by
\begin{eqnarray}\label{eq:3}
&&V(z_1,z_2,t; R_1, R_2) =
\nonumber \\
&& \sum_{i, \alpha=1}^2 \left(\frac{-Z_\alpha}{\sqrt{( z_i-R_\alpha)^2+a }}\right)
 +\frac{1}{\sqrt{( z_1-z_2)^2+b }}
 \nonumber \\
&&  +\frac{Z_1Z_2}{\sqrt{( R_1-R_2)^2+c }}+(z_1+z_2) \varepsilon(t)
\nonumber\\
\end{eqnarray}
where $ Z_1=1$ and $Z_2=1$ are the charges of nuclei and the screening parameters $a$, $b$ and $c$ are responsible for the softening of the electron-nucleus, electron-electron and nucleus-nucleus interactions, respectively. Values of these parameters are set to the same values as used in our previous works [23,29] and also the work of Camiolo et al. [30]. The instantaneous positions of the nuclei needed for the evaluation of the forces are calculated based on the Verlet algorithm [24,31]. The details of the numerical simulation is presented in [32].

The laser parameters in this work, $\lambda\ =\ 390\ \&\ 532$ nm wavelengths and $1\times10^{14}$, $5\times10^{14}$, $1\times10^{15}$ \&\ 5$\times10^{15}$ Wcm$^{-2}$ intensities, respectively correspond to the Keldysh parameter [33] values 2.51 \& 1.84, 1.12 \& 0.82, 0.79 \& 0.58, and 0.35 \& 0.26 for the PI-H$_2$ subsystem and 3.40 \& 2.49, 1.52 \& 1.11, 1.07 \& 0.79, and 0.48 \& 0.35 for the H$_2^+$ subsystem. The time-dependent induced dipole moments $\langle z\rangle$ are calculated based on the time-dependent wave function obtained via solution of the TDSE. For all cases, a pulse shape of $\varepsilon(t)=E_0 sin^2 (\frac{t\pi }{\tau}) cos(\omega_0 t)$  is used for the electric field of the laser pulse, and the pulse length is set to $\tau=$ 8 optical cycles (o.c.) along with a fixed time step of $\Delta t = 0.02$. The simulation box is the same as that introduced in detail in Ref. [29] in which $d_{H_2}$ = 10, $d_{H_2^+}$=500 and $d_{H_2^{2+}}$ = 580 (corresponding to the positions of the region boundaries at $\pm 5$, $\pm 250$ and $\pm 290$, respectively) for $\lambda$ = 390 nm wavelength, and to 10, 1200 and 1280 (corresponding to the positions of the region boundaries at $\pm 5$, $\pm 600$ and $\pm 640$, respectively) for $\lambda$ = 532 nm wavelength. 

The power spectrum of high harmonic generation (HHG) obtained from the mean value of the acceleration of each electron via [34]
\begin{eqnarray}
\label{eq:4}
S(\omega)=\begin{vmatrix}\frac{1}{T}\int_{0}^{T}\langle \ddot{r}(t)\rangle H(t)exp(-\begin{scriptsize}\end{scriptsize}i \omega t)dt \end{vmatrix}^2,
\end{eqnarray}
where
\begin{eqnarray}\label{eq:5}
  H(t)= \frac{1}{2}[1-cos(2\pi \frac{t}{T})],
\end{eqnarray}
is the Hanning window function and $T$ is the total pulse duration. This function reduces the effect of nondecaying components in the dipole acceleration that last after the laser pulse is switched off [35,36].
The time frequency profiles $W(\omega,t)$, TFP, of the high harmonics are obtained via a Morlet wavelet transform [34,37] of the time-dependent dipole acceleration given by
\begin{eqnarray}
\label{eq:6}
&&W(\omega,t)=
\sqrt{\frac{\omega}{\pi^{\frac{1}{2}}\sigma}}\times
\nonumber\\
&& \int_{0}^{\infty}\langle \ddot{r}(t')\rangle exp(-i \omega (t'-t))exp\left(\frac{-\omega^2 (t'-t)^2}{2\sigma ^2}\right) dt',
\nonumber \\
\end{eqnarray}
in which $\sigma$=2$\pi$.

In our previous work [24], we focused to the dynamics of individual \textit{homolytic} 
$(e_1 H_\alpha^+-H_\beta^+ e_2 )$ 
and \textit{ionic} $(H_\alpha^--H_\beta^+ )$ states formed transiently during the evolution of the two-electron wave packet of H$_2$ subsystem. In this work, we investigate the dynamics of two-electronic PI-H$_2$ and one-electronic H$_2^+$ subsystems and their contributions to the overall high harmonic generation spectrum of H$_2$. For this purpose, the overall HHG spectrum $S(\omega)$ in equation (4) can be decomposed as 
\begin{eqnarray}
\label{eq:7}
S(\omega) \simeq S_{H_2}(\omega) + S_{H_2^{+}}(\omega) + S_{H_2^{2+}}(\omega) + S_{Q}(\omega)
\end{eqnarray}
where the terms on the right hand side are the power spectrum of HHG arising from the PI-H$_2$, H$_2^{+}$, H$_2^{2+}$ and quasi-H$_2^{2+}$ subsystems, respectively. As recombination of electrons with nuclei are impossible in H$_2^{2+}$ and quasi-H$_2^{2+}$ regions, the contributions of the last two terms are negligible, and thus equation (7) is reduced to 

\begin{eqnarray}
\label{eq:8}
S(\omega) \simeq  S_{H_2}(\omega) + S_{H_2^{+}}(\omega).
\end{eqnarray}

$\vspace{-1cm}$

\section{Results and Discussion}

The evolution of the populations of the PI-H$_2$ and H$_2^{+}$ subsystems and the two regions of H$_2^+$ (H$_2^+$(I) and H$_2^+$(II)) are calculated and plotted in Fig. 2. It can be seen from this figure that at the lowest laser pulse intensity ($1 \times 10^{14}$ Wcm$^{-2}$), the PI-H$_2$ is the dominant subsystem, and the H$_2^+$ subsystem is populated slightly only in the last couple of cycles of the laser pulse. Therefore, it can be expected that the ionized H$_2^+$ subsystem does not considerably contribute to the HHG spectrum at this intensity. By increasing the intensity of the laser pulse, the population of the PI-H$_2$ subsystem drops sharply after initial cycles; however, the population of the H$_2^+$ subsystem begins to rise. 
Moreover, increasing intensity of the laser pulse shifts the PI-H$_2$ population depletion stage to shorter times. This intensity-dependence of the depletion time is similar for laser pulses of both $\lambda\ =$ 390 \&\ 532 nm wavelengths.

Fig. 2 shows also that the population of the H$_2^+$ subsystem rises, along with the decrease in the population of PI-H$_2$, up to a maximum and then is depleted to populate the Q-H$_2^{2+}$ and H$_2^{2+}$ subsystems. At the two highest intensities, the H$_2^+$ subsystem is re-populated at the last 1.5 cycles of the laser pulse due to the return of the wavepacket from the Q-H$_2^{2+}$ and H$_2^{2+}$ subsystems after the field amplitude is decreased. 

As it can be seen in Fig. 2, at $I=5\times10^{14}$ and $1\times 10^{15}$ Wcm$^{-2}$ intensities and during 4th and 5th cycles, the populations of the H$_2^+$(I) and H$_2^+$(II) regions follow weakly the direction of the laser field and oscillate alternatively, showing overall shaking of the wavepacket due to alternation of the direction of the field, but their oscillations decay very fast during the last few cycles. Time-variation of the H$_2^+$ subsystem population is a resultant of the variation of the population of the H$_2^+$(I) and H$_2^+$(II) regions.
At $I=5\times10^{15}$ Wcm$^{-2}$ intensity, population of the H$_2^+$(I) shows a sharp maximum during the 5th half cycle, while the population of H$_2^+$(II) decrease considerably during this half cycle. On the contrary, the population of H$_2^+$(II) shows a sharp maximum during the 4th half cycle, while the population of H$_2^+$(I) is approximately zero during this half cycle. Interestingly, the corresponding minimums and maximums and also the horizontal part of the H$_2^+$(I) and H$_2^+$(II) graphs do not overlap. 

It is possible to evaluate the time-dependent motion of the electron wavepacket belonging to each subsystem by calculating average position $\langle z\rangle$ of the electron in each corresponding region of the simulation box space and following their time variations.
The average positions of the PI-H$_2$ and H$_2^+$ subsystems are plotted in Fig. 3. This figure shows that at the lowest intensity ($1 \times 10^{14}$ Wcm$^{-2}$), the position of the PI-H$_2$ oscillates more or less symmetrically around $\langle z\rangle$ = 0 and follows the variation of the laser pulse. At higher intensities, variation of the position of this subsystem is complex, but still, oscillates around $\langle z\rangle$ = 0. However, variations of the position of the H$_2^+$ subsystem is not symmetric with respect to  $\langle z\rangle$ = 0 and is more complex. These asymmetric behaviors are in agreement with the results of Fig. 2 which indicate that the symmetry between the populations of H$_2^+$(I) and H$_2^+$(II) breaks down due to the intensity and other characteristics of the laser field. It can also be seen from Fig. 3 that at all intensities and wavelengths, the position expectation value of the H$_2^+$ subsystem oscillates out-of-phase with respect of the electric field of the laser pulse and has a delay with respect to the oscillations of the laser pulse.

The HHG spectra of the two regions of the H$_2^+$ subsystem, i.e. H$_2^+$(I) and H$_2^+$(II) (Fig. 1), and also the comparison between these regions and the HHG spectra of the H$_2^+$ subsystem are presented in Fig. 4. At the low intensities, the symmetry does not break down and H$_2^+$(I) and H$_2^+$(II) regions have similar HHG spectra almost over all harmonic orders. By increasing the intensity, the asymmetric behavior becomes more apparent until the HHG of H$_2^+$(I) overcomes that of H$_2^+$(II) for almost all harmonic orders at $5\times 10^{15}$ Wcm$^{-2}$ intensity. 
 
The HHG spectra obtained for PI-H$_2$ and H$_2^+$  subsystems and the whole H$_2$ system for
$\lambda$ = 390 \&\ 532 nm wavelengths and $I=1\times 10^{14}, 5\times10^{14}$, $1\times 10^{15}$ \&\ $5\times 10^{15}$ Wcm$^{-2}$ intensities are demonstrated in Fig. 5. 
According to this figure, at $1\times 10^{14}$ Wcm$^{-2}$ intensity, only the PI-H$_2$ subsystem takes part in high harmonic generation and the contribution of the H$_2^+$ subsystem is not significant. It is also clear from Figs. 7 and 8 that the time frequency profile of the generated high harmonics (TPF-HH) obtained for the PI-H$_2$ subsystem and the whole H$_2$ system are almost similar. 
For other intensities, the PI-H$_2$ and H$_2^+$ subsystems compete at low harmonic orders, while the H$_2^+$ subsystem dominates at high harmonic orders because of the higher I$_p$ of H$_2^+$ with respect to PI-H$_2$.

It can be deduced from Figs. 4 and 5 that the HHG spectra follows a non-perturbative plateau behavior immediately after a steep descent at small harmonic orders. This plateau region ends with a relatively sharp so-called cutoff.
The cutoff region can be explained by using the three-step model [39,40]. According to this model, after ionization, electrons accelerate in the laser field and can gain a kinetic energy with a maximum of 3.17 times of ponderomotive potential (U$_p$). So the maximum energy which can be released when one electron recombine with the parent ion (H$_2^+$ and H$_2^{2+}$), i.e. the cutoff energy E$_c$, can be evaluated by 
\begin{eqnarray}
\label{eq:9}
 E_c = I_p + 3.17 U_p
\end{eqnarray}

where $E_c$ and $I_p$ are respectively the cutoff energy and the ionization energy of the selected subsystem, and $U_p$ is the ponderomotive energy of the electron in the laser field. 
Electrons emitted in different periods traverse the nuclei more than once and can reach combined maximum return kinetic energies of 5.55$U_p$ = 3.17$U_p$ + 2.38$U_p$ and 4.70$U_p$ = 3.17$U_p$ + 1.53$U_p$ for first and third electronic return combined and first and second electronic return combined, respectively [27].
In Fig. 6, we focus on the HHG spectra produced by the PI-H$_2$ and H$_2^+$ subsystems at $\lambda$ = 390 \&\ 532 nm wavelengths and $I=1\times 10^{15}$ intensity. The two-electron cutoffs shown in this figure and their related new plateaus correspond to the first and second electronic return (1+2 NSDR) with maximum return kinetic energy of 4.70$U_p$. Moreover, the cutoff related to the first and third electronic return (1+3 NSDR) with maximum return kinetic energy of 5.55$U_p$ does not appear in the HHG spectra.
At $\lambda$ = 390 nm, the HHG spectrum of H$_2^+$ subsystem has only one plateau. At $\lambda$ = 532 nm, however, two plateau can be detected for this subsystem.
It is interesting that the second plateau for the PI-H$_2$ and H$_2^+$ subsystems are absent in other intensities investigated in Fig. 5.

According to Fig. 2, at $1\times 10^{15}$ Wcm$^{-2}$ intensity and $\lambda$ = 390 nm, the population of PI-H$_2$ subsystem becomes zero at the 5th cycle. Correspondingly, in Fig. 7, it can be seen that the intensities of the signals in the TFP-HH of the PI-H$_2$ subsystem decrease effectively after the 5th cycle. It is also shown in Fig. 7 that the intensities of the signals in the TFP-HH of the H$_2^+$ subsystem remain strong until the last cycle. Moreover, the TFP-HH of the overall H$_2$ system is similar to the TFP-HH of its H$_2^+$ subsystem after the 5th cycle. It can be concluded from these results that at $1\times 10^{15}$ Wcm$^{-2}$ intensity in the time interval between the 5th and 8th cycles, the H$_2^+$ subsystem contributes dominantly to the HHG. 

As it is showed in Fig. 5, at $5\times 10^{15}$ Wcm$^{-2}$ intensity, the contribution of the H$_2^+$ subsystem to the HHG is much more significant as compared to that of the PI-H$_2$ subsystem. According to Figs. 7 and 8, the intensity of signals in TFP-HH of the PI-H$_2$ subsystem are weak in comparison to those of the H$_2^+$ subsystem. Moreover, the contribution of the PI-H$_2$ subsystem to HHG is negligible after the 4th cycle at this intensity.

If the laser pulses comprise many optical cycles and have intensities such that non-dipole and relativistic effects can be neglected, the harmonic angular frequencies $\omega$ are only emitted at odd multiples of the laser angular frequency ($\omega\ = q\omega_0$, and q = 1, 3, 5, . . .) due to the inversion symmetry of the system in the field [38]. As discussed elaborately in Ref. [36], it seems that the complex patterns appearing in the HHG spectra of the H$_2^+$ system originated from two effects. The first effect is a non-adiabatic frequency red shift of the harmonics which can be seen for low harmonic orders and almost independent of the carrier envelope phase (CEP). The second effect comes from a spatially asymmetric emission along the z direction which breaks down the odd-harmonic rule. These effects also lead to complex patterns in the HHG spectra of the H$_2$ system obtained in this work.
At $1\times 10^{14}$ Wcm$^{-2}$ intensity and $\lambda$ = 390 nm wavelength, the odd-harmonic orders from 1 to 23 can be seen clearly in the HHG spectra of the H$_2$ system, but for $\lambda$ = 532 nm wavelength, some odd-harmonic orders can be distinguished (see Fig. 5). At higher intensities, the total HHG spectra shows an almost simple pattern at both low and high harmonic orders, but a complex pattern at the intermediate harmonic orders.  For instance, at $1\times 10^{15}$ Wcm$^{-2}$ intensity and $\lambda$ = 532 nm wavelength, the odd harmonic orders between 1 to 11 are evident. After the 11th harmonic order, the peaks become complex. Near cutoff region, between 50 to 72 orders, the spectrum becomes simple again, and interestingly, the even harmonics appear which is related to H$_2^+$ subsystem and resulted from constructive combination of the H$_2^+$(I) and H$_2^+$(II) regions. Comparingly, at $5\times 10^{15}$ Wcm$^{-2}$ intensity and $\lambda$ = 390 nm wavelength, the spectrum between 79 to 99 orders is simple, but the even harmonics replaced by the odd harmonics with red shift. 

As discussed in our previous work (see Fig. 3 of ref. [24]), the internuclear distance in initial cycles does not change significantly, while there is considerable population in the pre-ionized H$_2$ subsystem. So very similar variations and fall-offs are observed in the populations of the PI-H$_2$ and H$_2^+$ subsystems in the calculation with the freed nuclei.
If the dynamics of nuclei (classically) are taken into account, the HHG spectrum of the H$_2$ system is similar to the HHG spectrum of the H$_2$ system with fixed nuclei (Fig. 9). Interestingly, at $1\times 10^{15}$ Wcm$^{-2}$ intensity and $\lambda$ = 532 nm wavelength, the even harmonics between 50 to 72 orders are replaced by the odd harmonics with blue shift (see Fig. 10).

\section{Conclusion}
The HHG spectrum of the most simple two-electron system, hydrogen molecule, and contributions of its two-electron and one-electron subsystems (PI-H$_2$ and H$_2^{+}$) to the overall HHG spectrum were studied in this work. TDSE was solved numerically for a one-dimensional model of H$_2$ system in linearly polarized laser pulses of different intensities ($1\times10^{14}$, $5\times10^{14}$, $1\times10^{15}$ \&\ 5$\times10^{15}$  Wcm$^{-2}$) and wavelengths (390 \&\ 532 nm). The intensity and wavelength of the laser pulse affect the contributions of the subsystems to the HHG spectrum. At the lowest laser pulse intensity ($1 \times 10^{14}$ Wcm$^{-2}$), the pre-ionized H$_2$ (PI-H$_2$) is the dominant subsystem, and the ionized H$_2^+$ subsystem does not contribute to the HHG production. By increasing the intensity of the laser pulse, the contribution of H$_2^+$ subsystem becomes more significant. This effect can be discussed by comparing the intensities of the signals in the TFP-HH of PI-H$_2$ and H$_2^+$ subsystems. 
  
The population exchange among the main subsystems, i.e. PI-H$_2$ and H$_2^{+} $, and Q-H$_2^{2+}$ and H$_2^{2+}$ subsystems are also influenced by the intensity and wavelength of the laser pulse which are explained in section (IV). Contributions of two latter subsystems (Q-H$_2^{2+}$ and H$_2^{2+}$) to HHG spectrum is negligible.

The time-dependent evolution of the electron wave packet belonging to each subsystem was analyzed by calculating average position $\langle z\rangle$ of the electron in each corresponding region of space and following their time variations. Contrary to the PI-H$_2$ subsystem, variations of the position expectation value of the H$_2^+$ subsystem is not symmetric with respect to $\langle z\rangle$ = 0.
This is because of breaking down the symmetry between the populations of H$_2^+$(I) and H$_2^+$(II) regions of H$_2^+$ subsystem especially at high intensities.

We searched for the NSDR and SPEAR cutoffs in the HHG spectra. As the equilibrium internuclear distance (2.13 a.u.) is smaller than 8 a.u., we could not detect SPEAR. This agrees with the results which Madsen's group reported [27, 28].
At $\lambda$ = 390 \&\ 532 nm wavelengths and $I=1\times 10^{15}$ intensity, we detected the two-electron cutoffs corresponding to 1+2 NSDR with the maximum return kinetic energy of 4.70$U_p$, but the cutoffs related to 1+3 NSDR with the maximum return kinetic energy of 5.55$U_p$ did not appear in the HHG spectra. 

At $1\times 10^{14}$ Wcm$^{-2}$ intensity, the HHG spectra of the H$_2$ system is simple. At higher intensities, however, the total HHG spectra shows an almost simple pattern at both low and high harmonic orders, but a complex pattern at the intermediate harmonic orders.  
At $1\times 10^{15}$ Wcm$^{-2}$ intensity and $\lambda$ = 532 nm wavelength, the even harmonics appear near cutoff region, i.e between 50 to 72 orders. Comparingly, at $5\times 10^{15}$ Wcm$^{-2}$ intensity and $\lambda$ = 390 nm wavelength, the even harmonics near cutoff region, i.e between 79 to 99 orders, replaced by the odd harmonics with red shift. 

The HHG spectrum of the H$_2$ system with freed nuclei is similar to the HHG spectrum of the H$_2$ system with fixed nuclei. However, at $1\times 10^{15}$ Wcm$^{-2}$ intensity and $\lambda$ = 532 nm wavelength, the even harmonics between 50 to 72 orders are replaced by the odd harmonics with blue shift.

\begin{acknowledgments}
 
We acknowledge Tarbiat Modares University and University of Isfahan for financial supports and research facilities. 

\end{acknowledgments}


\section{References}
[1] T. Brabec and F. Krausz, Rev. Mod. Phys. 72, 545 (2000).

[2] M. Drescher, M. Hentschel, R. Kienberger, G. Tempea, C. Spielmann, G. A. Reider, P. B. Corkum, and F. Krausz, Science 291, 1923 (2001).

[3] P. Agostini and L. F. DiMauro, Rep. Prog. Phys. 67, 813(2004).

[4]  M. Lein, J. Phys. B 40, R135 (2007).

[5] A. D. Bandrauk, S. Barmaki, S. Chelkowski, and G. Lagmago, Kamta, in Progress in Ultrafast Intense Laser Science, edited by K. Yamaguchi et al. (Springer, New York, 2007), Vol. III, Chap. 9.

[6] J. P. Marangos, S. Baker, N. Kajumba, J. S. Robinson, J. W. G. Tisch, and R. Torres, Phys. Chem. Chem. Phys. 10,35 (2008), and references therein.

[7] F. Krausz and M. Ivanov, Rev. Mod. Phys. 81, 163 (2009).

[8] C. C. Chirila, I. Dreissigacker, E. V. V. D. Zwan, M. Lein, Phys. Rev. A 81, 033412 (2010)

[9] B. D. Bruner, Z. Masin, M. Negro, F. Morales, D. Brambila, M. Devetta, and S. Patchkovskii, Faraday discussions, 194, 369-405 (2016).

[10] K. C. Kulander, K. J. Schafer, and J. L. Krause, Atoms in Intense Laser Fields, edited by M. Gavrila (Academic, New York, 1992)

[11] J.L. Krause, K.J. Schafer, K.C. Kulander, Phys. Rev. Lett. 68, 3535 (1992).

[12] M. Brics, J. Rapp, and D. Bauer, Phys. Rev. A 93, 013404 (2016)

[13] C. B. Madsen, F. Anis, L. B. Madsen, and B. D. Esry, Phys. Rev. Lett. 109(16), 163003 (2012).

[14] R. E. F. Silva, F. Catoire, P. Riviere, H. Bachau, and F. Martin, Phys. Rev. Lett., 110(11), 113001 (2013).

[15] F. Catoire, R. E. F. Silva, P. Riviere, H. Bachau, and F. Martin, Phys. Rev. A, 89(2), 023415 (2014).

[16] B. Feuerstein, and U. Thumm, Phys. Rev. A, 67(4), 043405 (2003).

[17] S. Chattopadhyay, S. Bauch, and L. B. Madsen, Phys. Rev. A, 92(6), 063423 (2015).

[18] B. Y. Chang, S. Shin, A. Palacios, F. Martin, and I. R. Sola,  J. Phys. B, 48(4), 043001(2015).

[19] A. Kastner, F. Grossmann, R. Schmidt, J. M. Rost, Phys. Rev. A, 81, 023414 (2010)

[20] H. Ahmadi, M. Vafaee, and A. Maghari, J. Phys. B, 49(3), 035602 (2016).

[21] Y. C. Han, and L. B. Madsen, Phys. Rev. A, 87(4), 043404 (2013).

[22] M. Vafaee, H. Ahmadi, and A. Maghari, J. Phys. B, 50(2), 025601 (2016).

[23] B. Buzari, M. Vafaee, H. Sabzyan, Phys. Rev. A 85, 033407 (2012).

[24] B. Buzari, M. Vafaee, H. Sabzyan, J. Phys. B: At. Mol. Opt. Phys. 46 245401 (2013).

[25] M. Awasthi, Y. V. Vanne, A. Saenz, A. Castro, and P. Decleva, Phys. Rev. A, 77(6), 063403 (2008).

[26] P. Koval, F. Wilken, D. Bauer, and C. H. Keitel,  Phys. Rev. Lett. 98, 043904 (2007).

[27] K. K. Hansen and L. B. Madsen, Phys. Rev. A 93, 053427 (2016)

[28] K. K. Hansen and L. B. Madsen, Phys. Rev. A 96, 013401 (2017)

[29] M. Vafaee, F. Sami, B. Shokri, B. Buzari, and H. Sabzyan, J. Chem. Phys. 137, 044112 (2012).

[30] G. Camiolo, G. Castiglia, P. P. Corso, E. Fiordilino, and J. P. Marangos, Phys. Rev. A, 79, 063401 (2009)

[31] L. Verlet, Phys. Rev. 159, 98 (1967).

[32] F. Sami, M. Vafaee, and B. Shokri, J. Phys. B, 44, 165601 (2011).

[33] L.V. Keldysh, Zh. Eksp. Teor. Fiz. 47, 1945 (1964)[Sov. Phys. JETP 20, 1307 (1965)].

[34] A. D. Bandrauk, S. Chelkowski, H. Lu, Chem. Phys. 414, 73-83 (2013).

[35] S. Camp, K. J. Schafer, and M. B. Gaarde Phys. Rev. A 92, 013404 (2015).

[36] H. Ahmadi, M. Vafaee, and A. Maghari, Phys. Rev. A 94, 033415 (2016).

[37] C. Chandre, S. Wiggins, T. Uzer, Physica D 181 (2003) 171.

[38] C. J. Joachain, N. J. Kylstra, and R. M. Potvliege, Atoms in Intense Laser Fields (Cambridge University Press) (2012)

[39] P. B. Corkum, Phys. Rev. Lett., 71(13), 1994 (1993).

[40] M. Lewenstein, Ph. Balcou, M. Yu. Ivanov, Anne L'Huillier, and P. B. Corkum. Phys. Rev. A 49, 2117 (1994).


\bibliography{p7}
%


\end{document}